\documentclass[aps,prd,twocolumn,groupedaddress]{revtex4-1}
\usepackage{graphicx}
\include{amssym}

\begin{document}
\title{Isospin symmetry breaking and gluon anomaly}
\author{A.\,A. Osipov}
\email[]{aaosipov@jinr.ru}

\affiliation{Joint Institute for Nuclear Research, Bogoliubov Laboratory of Theoretical Physics, 141980 Dubna, Russia}

\begin{abstract}
The contribution of the electromagnetic interaction to the self-energy of pseudo-Goldstone bosons, as well as to the mixing angles and weak decay constants, is calculated in the first nonleading order in the simultaneous expansion in powers of $1/N_c$, momenta and quark masses. Particular attention is paid to the isospin symmetry breaking which is generated by means of $\eta$ and $\eta'$ admixtures to the pion. It is shown that the gluon anomaly enhances the electromagnetic contribution to the $\pi^0$-$\eta$ and $\pi^0$-$\eta'$ mixing angles, which facilitates the restoration of isospin symmetry in the spectrum of pseudo-Goldstone states at next-to-leading order. The contribution of the electromagnetic interaction to the $\eta$-$\eta'$ mixing angle is shown to be only about 3\%.
\end{abstract}

\maketitle

\vspace{0.5cm}


\section{Introduction}
Electromagnetic interactions play a special role in isospin symmetry breaking. Their contribution is comparable to the effect of the difference in the masses of the $u$- and $d$-quarks, so any attempt at a precise estimate of the quark masses requires careful consideration of electromagnetic effects \cite{Dashen:69,Weinberg:77,Gasser:82}. Since one of the main sources on quark masses is the phenomenological values of the masses of pseudoscalar mesons, the task of calculating the contribution of virtual photons to the self-energy of pseudoscalars becomes mandatory. It is not surprising that much research has been devoted to studying this problem, especially within the framework of traditional chiral perturbation theory \cite{Gasser:85,Donoghue:93,Urech:95,Neufeld:95,Neufeld:96,Baur:96,Urech:97,Moussallam:97, Knecht:00}. 

While the unequal quark masses $m_u\neq m_d$ and the electromagnetic interaction lead to the breaking of $SU(2)$  symmetry, the gluon anomaly produces the opposite effect, restoring it. The latter was demonstrated by Gross, Treiman and Wilczek \cite{Gross:79}, who used $SU(3)_L\times SU(3)_R$ current algebra and PCAC (Partially Conserved Axial-vector Current) techniques to study the consequences of the axial-vector $U(1)_A$ symmetry breaking on the spectrum and the mixing angles of pseudo-Goldstone bosons. The most noticeable effect of isospin symmetry breaking in the presence of a gluon anomaly that they found is the small value of the $\pi^0$-$\eta$ mixing angle $\epsilon\simeq 0.56^\circ$. This result occurs in the absence of electromagnetic interactions and under the assumption that the dynamical degrees of freedom of the $\eta'$ meson are frozen (there is no $\eta$-$\eta'$ mixing). This also follows from chiral perturbation theory at the lowest order \cite{Gasser:85}.

The aim of this work is to study the phenomenon of restoration of isospin symmetry beyond the leading order (LO), or more precisely, in the next-to-leading order (NLO). This is the minimum approximation within which one can not only obtain a satisfactory description of the pseudo-Goldstone nonet spectrum, but also go beyond Dashen's theorem \cite{Dashen:69}, which is known to be strongly violated in nature \cite{Donoghue:93,Prades:97}. 

Our consideration involves going beyond current algebra by using large $N_c$ chiral perturbation theory ($1/N_c$-ChPT) \cite{Leutwyler:96a,Leutwyler:98,Taron:97,Kaiser:00}, where $N_c$ is the number of quark colors, and, as a consequence, taking into account both the electromagnetic interactions and singlet-octet mixing. The latter, as is well known, has a significant impact on the magnitude of the angle $\epsilon$ already at LO, which in turn creates difficulties in describing the amplitude of the $\eta\to 3\pi$ decay. As Leutwyler argued in \cite{Leutwyler:96b}, such a discrepancy should be eliminated when going beyond the LO consideration. This is precisely what we observe. As will be shown below, the main role in the restoration of isospin symmetry at the next-to-leading order of $1/N_c$ expansion belongs to the enhancement mechanism of the electromagnetic contribution by the $U(1)_A$ anomaly.

The $1/N_c$-ChPT is a consistent framework for dealing with the nonet of pseudoscalar mesons. It is known that a theoretically correct description of the $\eta'$ physics is based on the $1/N_c$ expansion of QCD \cite{Hooft:74, Witten:79a}, which explains its large mass ($m_{\eta'}\simeq 1\,\mbox{GeV}$) by the contribution of the gluon $U(1)_A$ anomaly. In the limit $N_c\to\infty$ the gluon anomaly is suppressed. The chiral symmetry group is extended to $U(3)_L\times U(3)_R$ and as a result of its spontaneous breaking, nine Goldstone modes are excited. Only in this case does a coherent effective field theory exist, since all nine Goldstone bosons are treated as dynamical variables. This effective theory is none other than $1/N_c$-ChPT. Its vertices are classified by powers of momenta $p_\mu$, quark masses $m_q$ and $1/N_c$ in accordance with the counting rule $\mathcal O(p^2)=\mathcal O(m_q)=\mathcal O(1/N_c)$. It is this Lagrangian that we will use in our calculations  \cite{Leutwyler:96a,Taron:97}.

The Lagrangian describing the electromagnetic interactions of pseudoscalars is also well known \cite{Prades:97}. Its leading part contains two low-energy constants $C_1$ and $C_2$, and the next step of the chiral expansion includes 24 operators. It is significant that only ten of them have the order $\mathcal{O}(1/N_c)$, and only seven of these ten operators with low-energy constants $\tilde K_3$, $\tilde K_4$, $\tilde K_5$, $\tilde K_6$, $K_9$, $K_{10}$ and $K_{11}$ 
contribute to the masses of the pseudo-Goldstone states. These are the ones we will use.

The problem is further simplified by the fact that the masses and decay constants of $\pi$, $K$, $\eta$ and $\eta'$  mesons at next-to-leading order depend only on four linear combinations of the seven couplings: $k_1=2\tilde K_3-\tilde K_4$, $k_2=\tilde K_5+\tilde K_6$, $k_3=K_9+K_{10}$, and $k_4=K_{10}+K_{11}$, which allows us to fix them based on the phenomenological values of the masses $m_{\pi^+}$, $m_{K^+}$, $m_{K^0}$ and the pion decay constant $f_\pi$. Certainly, the analytical expressions for these physical quantities also include some other parameters associated with strong interactions. These are the quark masses $m_q$, the low-energy constants $B_0$, $L_5$, $L_8$, etc. To fix them we will use some recent results of lattice simulations \cite{FLAG:22} together with constraints on the quark mass ratios \cite{Gasser:85,Kaplan:86,Leutwyler:96c}, detailed in \cite{Osipov:24a,Osipov:24b}.

It should be noted that the large-$N_c$ off mass-shell expressions for the electromagnetic contribution to the self-energies of pseudoscalars were worked out in \cite{Prades:97}. Here we reproduce this result and go further by including in the analysis the important contribution of the $U(1)_A$ gluon anomaly, whose influence is known to be sizeable \cite{Witten:79a,Weinberg:75,Veneziano:80,Schechter:80,Ohta:80,Eides:81}. We show that it enhances the effects of isospin and $SU(3)$ symmetry breaking by electromagnetic interactions. It happens like this: Electromagnetic interactions cause mixing of singlet-octet components of neutral fields $\phi_0$-$\phi_3$-$\phi_8$, due to which the anomaly penetrates into the elements of the meson mass matrix $\tilde M^2_{08}$ and $\tilde M^2_{03}$, which ultimately leads to a sizable increase in electromagnetic contributions to the $\pi^0$-$\eta$ and $\pi^0$-$\eta'$ mixing angles. As a result, the combined effect of the explicit ($m_u\neq m_d$) and electromagnetic isospin symmetry violations largely compensates each other.

As further motivation for this study, we note that we are not aware of any work in which electromagnetic corrections to the mixing angles $\pi^0$-$\eta$-$\eta'$ have been calculated in the framework of the $1/N_c$-ChPT, although the steps have already been taken to study the $\eta$-$\eta'$ mixing up to next-to-next-to-leading-order \cite{Oller:15}. A similar analysis for the $\pi^0$-$\eta$ and $\pi^0$-$\eta'$ mixing angles is impossible without a careful consideration of isospin-violating electromagnetic interactions. We hope that the results obtained here will be useful in taking this step. 

The outline of the paper is as follows. In Section 2 we present the effective Lagrangian, which is then used to calculate the masses, decay constants and mixing angles of pseudoscalar mesons. The contribution of virtual photons to the self-energy of charged particles is also given there. The masses of $\pi^\pm$, $K^\pm$, and $K^0$ ($\bar K^0$) mesons, as well as the procedure of the renormalization used, are discussed in Section 3. In Section 4, the parameters of the theory are fixed. When considering neutral fields, it is necessary to take into account their mixing effects. This occurs in both the kinetic and mass parts of the free Lagrangian. The field redefinition caused by the off-diagonal terms of the kinetic part is discussed in detail in Section 5. The corresponding changes in the decay constants are analyzed in Section 6, where a singlet-octet basis is used for this purpose. The spectrum of neutral states and mixing angles are presented in Section 7. Here, the main result of the work, the mechanism of enhancement of the electromagnetic contribution via the $U(1)_A$ anomaly, is presented in detail. In Section 8 we summarize our results.


\section{Effective Lagrangian to order $\mathcal{O}(\delta) $}

For convenience of chiral counting, a single parameter $\delta$ is usually introduced, for which $\mathcal{O}(m_q)=\mathcal{O}(p^2)=\mathcal{O}(1/N_c)=\mathcal{O}(\delta)$. It is also assumed that the square of the electric charge is of order $\mathcal{O}(e^2)= \mathcal{O}(\delta)$ \cite{Urech:95,Neufeld:95, Neufeld:96}. Then the LO vertices are counted as $\mathcal{O}(1)$, the NLO vertices correspond to $\mathcal{O}(\delta)$, etc. As a result, the effective Lagrangian of strong and electromagnetic interactions of pseudoscalar mesons is given by a sum of operators with increasing powers of $\delta$   
\begin{equation}
\label{eff}
{\mathcal L}={\mathcal L}^{(0)}+{\mathcal L}^{(1)}+\ldots 
\end{equation}

The LO part of the  Lagrangian density reads
\cite{Leutwyler:96a,Pich:89}
\begin{eqnarray}
 \label{LO}
{\mathcal L}^{(0)}&=&\frac{F^2}{4} \langle d_\mu U d^\mu U^\dagger  
+  \chi U^\dagger+\chi^\dagger U \rangle -\frac{F^2}{2}M_0^2\phi_0^2  \nonumber\\
&-&\frac{1}{4}F^{\mu\nu}F_{\mu\nu}-\frac{1}{2}\xi (\partial^\mu A_\mu)^2+C\langle QUQU^\dagger\rangle. 
\end{eqnarray}
Here $U=\exp (i\phi)$, $\phi =\sum_{a=0,...,8}\lambda_a \phi_a$, and $\lambda_a$ are Hermitian $U(3)$ matrices. The brackets $\langle \ldots \rangle$ stand for the trace in the flavor space. $F={\mathcal O(\sqrt N_c)}$ is a weak decay constant of pseudo-Goldstone bosons in the chiral limit $m_q\to 0$. Matrix $\chi=2B_0m$, $m=\mbox{diag} (m_u,m_d,m_s)$, where low-energy constant $B_0=-\langle \bar qq\rangle /F^2$ is associated with the quark condensate. The term $\propto \phi_0^2$ in (\ref{LO}) stems from the QCD $U(1)_A$ anomaly with $M_0=\mathcal{O}(1/\sqrt{N_c})$ being the mass of the singlet state $\phi_0$ at $m_q=0$. $d_\mu$ is a covariant derivative, incorporating the couplings to the electromagnetic field $A_\mu$: $ d_\mu U=\partial_\mu U-iA_\mu [Q,U]$, where $Q=\frac{e}{3}\,\mbox{diag}(2,-1,-1)$ is the diagonal matrix of quark charges. $F_{\mu\nu}$ is the electromagnetic field strength tensor $F_{\mu\nu}=\partial_\mu A_\nu - \partial_\nu A_\mu$. The gauge fixing parameter $\xi$ is henceforth taken to be equal to unity (Feynman gauge). The dimensional constant $C=\mathcal{O}(N_c)$. 

The last term in (\ref{LO}) leads to the electromagnetic contribution to the self-energies of $\phi^\pm =(\pi^\pm,K^\pm)$  mesons in full agreement with Dashen's theorem \cite{Dashen:69}
\begin{equation}
\label{Dashen}
    (\mu^2_{\phi^\pm})_{em}=2e^2\frac{C}{F^2}.
\end{equation}
This result is obtained by setting $\phi =\phi_{ph}/f_{\phi}$ in $U$. For large $N_c$, $f_{\phi}=F$.  

 The quark masses $m_q$ and the quark condensate $B_0$ in QCD must be renormalized. As a result, both quantities depend on the running renormalization-group scale $\mu_{\mbox{\tiny QCD}}$. A change in scale modifies each of them according to $m_q\to Z_M^{-1}m_q$, $B_0\to Z_M B_0$, while their product $\chi$ is an invariant quantity.  

Let us turn now to the Lagrangian ${\mathcal L}^{(1)}$ that accounts for the next-to-leading order corrections to ${\mathcal L}^{(0)}$ 
\begin{eqnarray}
\label{L1}
{\mathcal L}^{(1)}&=& L_5\langle d_\mu U^\dagger d^\mu U
                      (\chi^\dagger U+U^\dagger \chi )\rangle
                      \nonumber \\
  &+&L_8\langle \chi^\dagger U\chi^\dagger U+ h.c. \rangle
      +\frac{1}{2}\Lambda_1 F^2 \partial_\mu\phi_0\partial^\mu\phi_0
      \nonumber \\
  &+&\frac{i \Lambda_2}{2\sqrt{6}} F^2 \phi_0 \langle \chi^\dagger
      U-U^\dagger \chi\rangle  \nonumber \\
      &+& \tilde K_3F^2\langle QU^\dagger d_\mu UQd^\mu U^\dagger U + QUd_\mu U^\dagger Q d^\mu U U^\dagger\rangle \nonumber \\
   &+&\tilde K_4 F^2 \langle QU^\dagger d_\mu U Qd^\mu UU^\dagger \rangle \nonumber \\
   &+& \tilde K_5 F^2 \langle (d_\mu U^\dagger d^\mu U +d_\mu U d^\mu U^\dagger )Q^2\rangle \nonumber \\
   &+& \tilde K_6 F^2 \langle d^\mu U^\dagger d_\mu U Q U^\dagger Q U +d^\mu U d_\mu U^\dagger QUQU^\dagger\rangle \nonumber \\
   &+& K_9 F^2 \langle (\chi^\dagger U+U^\dagger\chi +\chi U^\dagger +U\chi^\dagger )Q^2\rangle +K_{10}F^2 \nonumber \\
   &\times& \langle(\chi^\dagger U+U^\dagger\chi)QU^\dagger QU+(\chi U^\dagger +U\chi^\dagger)QUQU^\dagger\rangle \nonumber \\
   &+&K_{11} F^2 \langle(\chi^\dagger U-U^\dagger\chi)QU^\dagger QU
   \nonumber \\
   &+&(\chi U^\dagger -U\chi^\dagger)QUQU^\dagger\rangle  +\ldots ,
\end{eqnarray}  
where only the terms relevant to this work are included. 

The QCD part of this Lagrangian contains four dimensionless low-energy constants (LECs): $L_5$, $L_8={\mathcal O}(N_c)$, $\Lambda_1$, $\Lambda_2={\mathcal O}(1/N_c)$ \cite{Leutwyler:96a}. The chiral logarithms arising from the calculation of one-loop meson diagrams constructed on the basis of the Lagrangian ${\mathcal L}^{(0)}$ are of the higher order $m_q/N_c\ln m_q={\mathcal O}(\delta^2)$ and can therefore be neglected. As a consequence, the coupling constants $L_5$ and $L_8$ are scale independent. 

In contrast, couplings $\Lambda_1$ and $\Lambda_2$ as well as the singlet field $\phi_0$ and $M_0$ depend on the QCD running scale $\mu_{\mbox{\tiny QCD}}$. This is because the singlet axial-vector current has nonvanishing anomalous dimension \cite{Kodaira:80,Espriu:82}. In the framework of $\delta$-expansion, the point was elaborated in \cite{Leutwyler:98,Kaiser:98}, where in particular scaling laws for the effective coupling constants were clarified
\begin{eqnarray}
M_0^2&\to&Z_A^2 M_0^2, \quad \phi_0\to Z_A^{-1}\phi_0, \nonumber \\
1+\Lambda_1&\to&Z_A^2(1+\Lambda_1), \quad 1+\Lambda_2\to Z_A(1+\Lambda_2).
\end{eqnarray}
Since the renormalization of the axial-vector current is subleading in $1/N_c$, such dependence appears first at the level of ${\mathcal L}^{(1)}$, so that $Z_A=1+\delta Z_A$, $\delta Z_A=\mathcal{O}(1/N_c)$. It follows that $2\Lambda_2 -\Lambda_1$ and $(1-\Lambda_1)M^2_0$ are the scale invariant combinations. The above relations are sufficient to verify the scale invariance of physical quantities obtained on the basis of the $\delta$-expansion.

The one-loop diagrams with virtual photons generated by the LO operators are of order $\mathcal{O}(\delta)$ and therefore must be taken into account. They suffer from the ultraviolet divergences. The seven LECs $\tilde K_i$ and $K_i$ in (\ref{L1}) are the needed counterterms to regularize them.  There are two such diagrams (see Fig.\,1). In the dimensional regularization, only diagram (a) makes a nonzero contribution. It describes the $\phi^\pm\to\phi^\pm$ amplitude, which in the Feynman gauge for the photon propagator is given by
\begin{eqnarray}
\label{loops}
\mathcal{M}^{(a)}_{\phi^\pm}(p^2)&=&2e^2 \mu_{\phi^\pm}^2 \left[\lambda (\mu)+\frac{1}{32\pi^2} \left(\ln\frac{\mu^2_{\phi^\pm}}{\mu^2}-2\right)\right] \nonumber \\
&+& 4e^2 p^2 \left[\lambda (\mu)+\frac{1}{32\pi^2} \left(\ln\frac{\mu^2_{\phi^\pm}}{\mu^2}-1\right)\right] \nonumber \\
&+&\frac{e^2}{8\pi^2 p^2} \left(p^4-\mu^4_{\phi^\pm}\right)\ln\left(1-\frac{p^2}{\mu^2_{\phi^\pm}}\right).
\end{eqnarray}

Here and below we neglect terms quadratic in isospin breaking, i.e., $\sim e^2(m_d-m_u)$, $e^4$, $(m_d-m_u)^2$. It is for this reason that the electromagnetic contribution is missing in the LO expressions for the squared masses of the charged states $\phi^\pm$ in (\ref{loops}), where $\mu_{\pi^\pm}^2=B_0(m_u+m_d)$, $\mu_{K^\pm}^2=B_0(m_u+m_s)$. In the $\overline{MS}$ scheme \cite{Gasser:85}, the pole divergence at $d=2\omega=4$ is separated in a form
\begin{equation}
    \lambda (\mu) =\frac{\mu^{d-4}}{32\pi^2}\left(\frac{1}{\omega-2}-\psi (1)-\ln(4\pi)-1\right),  
\end{equation}
where $\mu$ is the scale on which the renormalization is performed, $\psi(1)=-\gamma_E$, and $\gamma_E$ is Euler's constant.  

\begin{figure}
\includegraphics[width=0.35\textwidth]{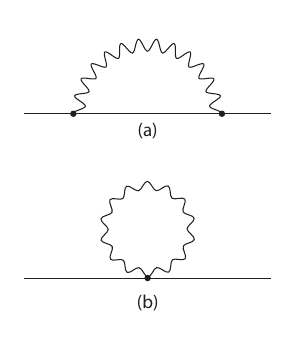}
\caption{Photon loop contribution to the self-energy of $\pi^\pm$ and $K^\pm$ mesons. Diagram (a) corresponds to formula (\ref{loops}). Diagram (b) is zero in dimensional regularization.}    
\label{fig1}      
\end{figure}


\section{Decay constants and masses of charged mesons}

NLO corrections contribute to both the decay constants and the masses of pseudoscalars. To find the decay constants we relate the bare fields $\phi$ to the physical states $\phi_{ph}$. In the case of charged particles, the transition to physical fields is accomplished by bringing the kinetic part of the Lagrangian to a canonical form, which is achieved by a corresponding redefinition of the original field
\begin{equation}
\label{fphi}
\phi =f_{\phi_{st}}^{-1}\,Z_{\phi}^{-\frac{1}{2}}\,\phi_{ph} =f_\phi^{-1} \phi_{ph},
\end{equation}
where $\phi_{ph} =(\pi^\pm, K^\pm, K^0, \bar K^0 )$, $f_{\phi_{st}}$ is the pure QCD part of the decay constant $f_\phi$.  The electromagnetic contribution is associated with the factor $Z_{\phi}$.

The NLO result for coupling $f_{\phi_{st}}$ is well known \cite{Goity:02}. Therefore, we proceed to calculate the electromagnetic contribution. The easiest way to do this is to redefine the field $\phi=f^{-1}_{st}\phi_R$. After this redefinition, the kinetic part of the Lagrangian $(\partial\phi_R)^2$ receives only electromagnetic corrections, which allows us to find $Z_{\phi}$.

Let us consider the $\phi_R^\pm\to\phi_R^\pm$ transition amplitude. To next-to-leading order in $\delta$ it is given by the following expression 
\begin{eqnarray}
\Pi_{\phi^\pm_R}(p^2)&=&p^2-\mu^2_{\phi^\pm}\left[1+ \frac{8\mu^2_{\phi^\pm}}{F^2}(2L_8-L_5)\right]\nonumber \\
&-&\frac{2e^2}{F^2}C\left(1-\frac{8 \mu^2_{\phi^\pm}}{F^2}L_5\right)+\mathcal{M}^{(a)}_{\phi^\pm}(p^2) \nonumber \\
&+& C^{(1)}_{\phi^\pm}+C^{(2)}_{\phi^\pm}p^2 =p^2-m^2_{\phi^\pm} (p^2),
\end{eqnarray}
where the coefficients $C^{(1)}_{\phi^\pm}$, $C^{(2)}_{\phi^\pm}$ represent the contributions of the contact terms contained in (\ref{L1}). For them we have
\begin{eqnarray}
C^{(1)}_{\pi^\pm}&=& -\frac{8}{9}e^2 B_0 \left[k_3(m_d+4m_u)+18\hat m k_4 \right] \nonumber \\
&\simeq& -\frac{4}{9}e^2 \mu^2_{\pi^\pm} \left(5k_3 +18 k_4 \right), \nonumber \\
C^{(1)}_{K^\pm}&=& -\frac{8}{9}e^2 B_0 \left[k_3(m_s+4m_u)+9(m_s+m_u) k_4 \right] \nonumber \\
&\simeq& -\frac{8}{9}e^2 B_0 \left[k_3(m_s+4\hat m)+9(m_s+\hat m) k_4 \right], \nonumber \\
C^{(2)}_{\pi^\pm}&=&C^{(2)}_{K^\pm}=\frac{4}{9}e^2\left(-2k_1+5k_2\right). 
\end{eqnarray} 
Since we take into account only the leading contribution in isospin symmetry breaking, in all expressions proportional to $e^2$ we replace $m_u, m_d\to \hat m=(m_u+m_d)/2$, e.g. $\mu^2_{K^\pm}\to\hat \mu^2_{K}=B_0(m_s+\hat m)$. 

The amplitude $\Pi_{\phi^\pm_R}(p^2)$ can be expanded in a series in the neighbourhood of the mass shell $p^2=m^2_{\phi^\pm}$ \cite{Osipov:92,Osipov:96}
\begin{eqnarray}
\Pi_{\phi^\pm_R}(p^2)&=&(p^2-m^2_{\phi^\pm})\left[1-\left(m^2_{\phi^\pm}(p^2)\right)'_{p^2=m^2_{\phi^\pm}}\right]+\ldots \nonumber \\
&=&(p^2-m^2_{\phi^\pm})Z_{\phi^\pm}+\mathcal O \left((p^2-m^2_{\phi^\pm})^2\right). 
\end{eqnarray}

This yields
\begin{eqnarray}
m^2_{\phi^\pm}&=&\mu^2_{\phi^\pm}\left\{1+ \frac{8 \mu^2_{\phi^\pm}}{F^2} (2L_8-L_5) - C^{(2)}_{\phi^\pm} \right. \nonumber \\
&-&\left. 2e^2\left[3\lambda(\mu)+\frac{1}{32\pi^2}\left(3\ln \frac{\mu^2_{\phi^\pm}}{\mu^2}-4\right) \right]\right\} \nonumber \\
&+&\frac{2e^2}{F^2}C\left(1- \frac{8 \mu^2_{\phi^\pm}}{F^2} L_5 \right)-C^{(1)}_{\phi^\pm},
\end{eqnarray}
and
\begin{eqnarray}
Z_{\phi^\pm}&=&1+C^{(2)}_{\phi^\pm}+\frac{e^2}{4\pi^2} \ln\left(1-\frac{m^2_{\phi^\pm}}{\mu^2_{\phi^\pm}}\right) \nonumber \\
&+&4e^2 \left[ \lambda(\mu ) + \frac{1}{32\pi^2} \left(\ln \frac{\mu^2_{\phi^\pm}}{\mu^2}+1\right)\right]. 
\end{eqnarray}

Given the approximations we use, the difference in the first logarithm is
\begin{equation}
1-\frac{m^2_{\phi^\pm}}{\mu^2_{\phi^\pm}}=- 8 \frac{\mu^2_{\phi^\pm}}{F^2} (2L_8-L_5)+\ldots,
\end{equation}
where, as in (\ref{loops}), we neglect the $\mathcal{O}(e^2)$ terms under the logarithm since they lead to a higher order effects in isospin symmetry breaking.

The self-energy function of neutral kaons does not receive electromagnetic contributions from the LO part of the effective Lagrangian
\begin{eqnarray}
\label{seK0}
m^2_{K^0}(p^2)&=&\mu^2_{K^0}\left[1+\frac{8 \mu^2_{K^0}}{F^2}(2L_8-L_5)\right]\nonumber \\
&-& C^{(1)}_{K^0}-C^{(2)}_{K^0}p^2,
\end{eqnarray}
where $\mu^2_{K^0}=B_0 (m_d+m_s)$, and 
 \begin{eqnarray}
C^{(1)}_{K^0}&=& -\frac{8}{9}e^2 B_0 (m_s+m_d) k_3 \simeq -\frac{8}{9}e^2 \hat \mu^2_{K} k_3, \nonumber \\
C^{(2)}_{K^0}&=&\frac{4}{9}e^2 (k_1+2k_2).
\end{eqnarray} 

From (\ref{seK0}) one finds
\begin{eqnarray}
m^2_{K^0}&=&\mu^2_{K^0}\left[1+ \frac{8 \mu^2_{K^0}}{F^2} (2L_8-L_5)\right] \nonumber \\
&-&C^{(1)}_{K^0}-\hat \mu^2_{K}C^{(2)}_{K^0}, \\
Z_{K^0}&=&1+C^{(2)}_{K^0}+\mathcal{O}(\delta^2).
\end{eqnarray}

To eliminate the pole singularities of the constants $Z_{\phi^\pm}$ and the masses $m_{\phi^\pm}$, one should redefine the low-energy constants $\tilde K_i, K_i$ in such a way that the renormalized finite parts $\tilde K_i^r(\mu), K_i^r(\mu )$ depend on scale $\mu$ and are given by the formulas
\begin{equation}
\tilde K_i=\Sigma_i\lambda (\mu)+\tilde K_i^r(\mu )
\end{equation}
(hereinafter we give expressions only for the running constants $\tilde K_i^r(\mu)$, the LECs $K_i^r(\mu )$ satisfy similar relations).

Since $\tilde K_i$ ($K_i$) does not depend on $\mu$, one has 
\begin{equation}
\mu\frac{d}{d\mu} \tilde K_i^r=-\frac{\mu^{d-4}}{16\pi^2}\Sigma_i.
\end{equation}
The solution of this equation relates the values of the finite constants at two arbitrary renormalization points $\mu$ and $\mu_0$
\begin{equation}
\tilde K_i^r(\mu)=\tilde K_i^r(\mu_0)+\frac{\Sigma_i}{32\pi^2}\ln\frac{\mu_0^2}{\mu^2}.
\end{equation}

The values of the $\beta$-functions $\Sigma_i$ are determined based on conditions that ensure the elimination of the pole singularities of the self-energies $m_{\phi^\pm}(p^2)$. This gives 
\begin{eqnarray}
\label{Sigmas}
&& 2\Sigma_3-\Sigma_4=2, \quad \Sigma_5+\Sigma_6=-1, \nonumber \\
&& \Sigma_9+\Sigma_{10}=0, \quad \Sigma_{10}+\Sigma_{11}=\frac{1}{4}.
\end{eqnarray}
 
At this point, it is useful to provide explicit expressions of the masses and decay constants at NLO, including electromagnetic corrections. For the decay constants the above expressions lead to 
\begin{equation}
\label{fphi}
f_\phi =f_{\phi_{st}} Z_{\phi}^{\frac{1}{2}},
\end{equation}
where, depending on the specific field $\phi_{ph}$, we got for $Z_{\phi}^{\frac{1}{2}}$ 
\begin{eqnarray}
\label{Z}
    Z^{\frac{1}{2}}_{\pi^\pm}&=& 1+\frac{2}{9}e^2\left(5k_2^r-2k_1^r\right) \nonumber \\
    &+&\frac{e^2}{16\pi^2}\left[\ln\frac{\mu_{\pi^\pm}^2}{\mu^2}+1+2\ln \left(- 8 \frac{\mu^2_{\pi^\pm}}{F^2} (2L_8-L_5)\right)\right], \nonumber \\
    Z^{\frac{1}{2}}_{K^\pm}&=& 1+\frac{2}{9}e^2\left(5k_2^r-2k_1^r\right) \nonumber \\
    &+&\frac{e^2}{16\pi^2}\left[\ln\frac{\hat\mu_{K}^2}{\mu^2}+1 +2\ln\left(- 8 \frac{\hat\mu^2_{K}}{F^2} (2L_8-L_5)\right)\right], \nonumber \\
    Z^{\frac{1}{2}}_{K^0}&=& 1+\frac{2}{9}e^2\left(k_1^r+2k_2^r\right).
\end{eqnarray}
Here $k_1^r=2\tilde K_3^r-\tilde K_4^r$ and $k_2^r=\tilde K_5^r+\tilde K_6^r$. It follows from (\ref{Sigmas}) that the sum $k_1^r+2k^r_2$ is scale-independent.

After renormalization, the expressions for the masses of pseudo-Goldstone bosons take the following form
\begin{eqnarray}
\label{pionmass}
    m_{\pi^\pm}^2&=&\mu^2_{\pi^\pm}\left(1+8 \frac{\mu^2_{\pi^\pm}}{F^2} (2L_8-L_5)\right) \nonumber \\
    &+&2e^2\frac{C}{F^2}\left(1-\frac{8 \mu^2_{\pi^\pm}}{F^2}L_5\right) \nonumber \\
    &-&\frac{e^2}{16\pi^2}\mu^2_{\pi^\pm}\left(3\ln \frac{\mu^2_{\pi^\pm}}{\mu^2}-4\right)\nonumber \\
    &-&\frac{4}{9}e^2 \mu^2_{\pi^\pm}(5k_2^r-2k_1^r-5k_3^r-18k_4^r). \\
\label{kaonmass}
    m_{K^\pm}^2&=&\mu^2_{K^\pm}\left(1+8\frac{\mu^2_{K^\pm}}{F^2} (2L_8-L_5) \right)  \nonumber \\
    &+&2e^2\frac{C}{F^2}\left(1-\frac{8\hat \mu^2_{K}}{F^2} L_5\right)                             \nonumber \\
    &-&\frac{e^2}{16\pi^2}\hat \mu^2_{K}\left(3\ln \frac{\hat\mu^2_{K}}{\mu^2}-4\right)   \nonumber  \\
    &-&\frac{4}{9}e^2 \left[\hat \mu^2_{K}(5k_2^r-2k_1^r-2k_3^r-18k_4^r) \right. \nonumber \\
    &-&\left. 3\mu^2_{\pi^\pm}k^r_3 \right].  \\
    \label{nkaonmass}
    m_{K^0}^2&=&\mu^2_{K^0}\left(1+8\frac{\mu^2_{K^0}}{F^2} (2L_8-L_5) \right) 
    \nonumber \\
    &-&\frac{4}{9}e^2\hat\mu^2_{K}\left(k_1^r+2k_2^r-2k_3^r \right).
\end{eqnarray}
Here  $ k_3^r=K_9^r+K_{10}^r$, and $k_4^r=K_{10}^r+K_{11}^r$.

Note that the corrections to the pseudoscalar meson masses due to electromagnetic interaction were calculated in \cite{Prades:97}. The formulas (\ref{pionmass}), (\ref{kaonmass}) and (\ref{nkaonmass}) are consistent with these estimates. The only minor difference is that when parameterizing the physical masses and decay constants of pseudo-Goldstone states, we use LO expressions for the meson masses. This is because these formulas previously formed the basis for estimating the quark masses \cite{Osipov:24a,Osipov:24b}, which we will rely on further. As shown in \cite{Oller:15}, differences in the analytical form of the expressions only manifest themselves at the NNLO level.

\section{Fixing parameters}

The formulas given above contain twelve parameters: $B_0$, $m_u$, $m_d$, $m_s$, $F$, $C$, 
\begin{equation}
\label{D}
\Delta\equiv 8B_0(2L_8-L_5)/F^2, \quad  \Delta'\equiv 8B_0L_5/F^2, 
\end{equation}
$k^r_1$, $k^r_2$, $k^r_3$, $k^r_4$. To make numerical estimates, we need to fix them.

Five quantities: $B_0$, the quark masses, and $\Delta$ were established in \cite{Osipov:24a,Osipov:24b} based on an analysis of the QCD mass formulas for $\pi^\pm$, $K^\pm$, and $K^0$ mesons in the NLO approximation with the use of additional information obtained on the lattice \cite{FLAG:22}: 
\begin{eqnarray}
B_0&=&2.682(53)\, \mbox{GeV} \quad \mbox{\cite{Borsanyi:2013}}, \\
\Delta&=&-0.57(5) \,\mbox{GeV}^{-1}, \\
m_u&=&2.14(7)\, \mbox{MeV}, \\
m_d&=&4.70(12) \,\mbox{MeV}, \\
m_s&=&93.13(2.25)\, \mbox{MeV}.
\end{eqnarray} 
All LECs correspond to the scale $\mu_{\mbox{\tiny QCD}}=2\,\mbox{GeV}$ in the $\overline{MS}$ subtraction scheme. Note that the above quark masses, within the specified errors, coincide with the latest data quoted by PDG \cite{PDG:24}: $m_u=2.16(4) \,\mbox{MeV}$, $m_d=4.70(4) \,\mbox{MeV}$, $m_s=93.5(5) \,\mbox{MeV}$. 

The constant $C$ can be determined from the spectral functions \cite{Das:67} if we additionally use the Weinberg sum rules \cite{Weinberg:67}
\begin{equation}
    C=\frac{3}{32\pi^2}m_\rho^2f_\rho^2\ln\frac{f^2_\rho}{f^2_\rho-f^2_\pi}=59.4 \times 10^{-6}\,\mbox{GeV}^4,
\end{equation}
where the following values of the mass and decay constant of the $\rho$-meson were used for the numerical estimate: $m_\rho=770\,\mbox{MeV}$, $f_\rho=154 \,\mbox{MeV}$ \cite{Ecker:89}.

To find the value of $\Delta'(\mu_{\mbox{\tiny QCD}}\!=\!2\, \mbox{GeV})$, we consider the ratio 
\begin{equation}
\label{fkpi}
    \frac{f_{K^+}}{f_{\pi^+}}=1\!+\!\frac{1}{2}(m_s\!-\!m_d)\Delta'
    \!+\!\frac{3e^2}{16\pi^2}\ln \frac{\hat\mu^2_{K}}{\mu^2_{\pi^+}},
\end{equation}
which is a direct consequence of (\ref{fphi}) and (\ref{Z}). Using the lattice result $f_{K^+}/f_{\pi^+}=1.1932(21)$ $(N_f\!=\!2\!+\!1\!+\!1 )$ \cite{FLAG:22}, we obtain
\begin{equation}
     \Delta'(\mu_{\mbox{\tiny QCD}}\!=\!2\, \mbox{GeV})=4.26(15) \,\mbox{GeV}^{-1}.
\end{equation}

Based on $\Delta$ and $\Delta'$, the values of coupling constants $L_5$ and $L_8$ can be determined. To do this, we must first to fix $F$, which in ChPT is estimated as $F\simeq (86\pm 10)\,\mbox{MeV}$ \cite{Prades:97}. In the following we will use the value $F=93.5\,\mbox{MeV}$, which practically coincides with the value $F=93.3\,\mbox{MeV}$ used in \cite{Urech:95} when studying the contribution of virtual photons in the $SU(3)$ approach. An additional argument in favor of the choice made can be obtained by calculating, based on the numerical values of $B_0$ and $F$, the magnitude of the quark condensate $|\langle\bar qq\rangle|^{1/3}=286\,\mbox{MeV}$. This is entirely consistent with FLAG's estimate:
 $|\langle\bar qq\rangle|^{1/3}=286\pm 23\,\mbox{MeV}$ \cite{FLAG:22}. As a result, we find: $L_5=1.74(2)\times 10^{-3}$, $L_8=0.75(2)\times 10^{-3}$, which is in excellent agreement with the estimates $L_5^r=1.4(5)\times 10^{-3}$, $L_8^r=0.9(3)\times 10^{-3}$ \cite{Prades:97}, or $L_5^r=2.2(5)\times 10^{-3}$, $L_8^r=1.1(3)\times 10^{-3}$ \cite{Gasser:85}, obtained in the standard ChPT.  

The value of $F$ that we use in some sense restrains the contribution of chiral logarithms of electromagnetic origin. This can be verified by examining the expression for the difference in the masses of charged and neutral pions in the leading approximation of the $\delta$-expansion
\begin{equation}
  m_{\pi^\pm}-m_{\pi^0}=\frac{e^2 C}{F^2 \mu_{\pi^\pm}}+\mathcal{O}(\delta^{3/2}), 
\end{equation}
from which it follows that for the given values of the parameters $m_{\pi^\pm}-m_{\pi^0}=4.7\,\mbox{MeV}$. This estimate is close to the experimental one, $(m_{\pi^\pm}-m_{\pi^0})_{\mbox{\footnotesize exp}}=4.6\, \mbox{MeV}$, and leaves only a 2\% window for NLO corrections.

To fix the four coupling constants $k^r_1$, $k^r_2$, $k^r_3$, $k^r_4$, we use the NLO result
\begin{eqnarray}
    f_\pi&=&F\left\{1+\hat m\Delta' +\frac{2}{9}e^2\left(5k_2^r-2k_1^r\right)\right. \nonumber \\
    &+&\left.\frac{e^2}{16\pi^2}\left[1+\ln\frac{\mu_{\pi^\pm}^2}{\mu^2}+2\ln(-2\hat m\Delta)\right]\right\},
\end{eqnarray}
and formulas (\ref{pionmass}), (\ref{kaonmass}), (\ref{nkaonmass}). Solving the system of four linear equations for $\mu=1\,\mbox{GeV}$, and using the experimental values of the meson masses $m_{\pi^\pm}$, $m_{K^\pm}$, $m_{K^0}$ and $f_{\pi^\pm}=92.277(95)\, \mbox{MeV}$ we find:
\begin{eqnarray}
\label{lin}
    k_1^r&=&0.114(12),\qquad\! k_2^r=-0.145(6), \nonumber\\  
    k_3^r&=&-0.088(1), \quad k_4^r=-0.041(3).
\end{eqnarray}
The error bars indicated in (\ref{lin}) are due mainly to the accuracy with which the constant $f_\pi$ is known. The values of these parameters remain within the specified uncertainties when $\mu$ is varied in the range $m_\rho<\mu<2\,\mbox{GeV}$. Since values (\ref{lin}) are based on formulas obtained in the NLO approximation, they cannot be directly compared with the results of \cite{Prades:97}, where these LECs were calculated to all orders in the standard ChPT at large $N_c$. 

Another question arises that requires some clarification. When fixing the constants $B_0$, $m_q$ and $\Delta$ in \cite{Osipov:24a,Osipov:24b}, the phenomenological masses of the $\pi^\pm$, $K^\pm$ and $K^0$ mesons were also used. In this case, however, the electromagnetic part of the self-energy was estimated empirically, i.e., without systematically calculating the contribution of virtual photons. Instead, two constants $\Delta^2_{em}$ and $\tilde\Delta^2_{em}$ were involved 
\begin{eqnarray}
   &&(m_{\pi^\pm}^2-m^2_{\pi^0})_{em}=\Delta^2_{em}, \\
   &&(m_{K^\pm}^2-m^2_{K^0})_{em}=\tilde\Delta^2_{em}.
\end{eqnarray}
Since the difference in masses of charged and neutral pions is mainly of electromagnetic origin, the first constant $\Delta^2_{em}$ can be estimated from the experimental values of the masses of these particles. The QCD contribution here is known to be insignificant $\sim (m_d-m_u)^2$ and is estimated as $(m_{\pi^\pm}-m_{\pi^0})_{\mbox{\tiny QCD}}=0.17(3)\, \mbox{MeV}$ \cite{Gasser:85}, which ultimately gives
\begin{equation}
    \label{empion}
   \Delta^2_{em}=1.21(1)\times 10^{-3}\,\mbox{GeV}^2.
\end{equation}

The second constant $\tilde\Delta^2_{em}$ is related to the violation of Dashen's theorem: 
\begin{equation}
(m_{K^\pm}^2-m^2_{K^0})_{em}=(1+\varepsilon )(m_{\pi^\pm}^2-m^2_{\pi^0})_{em}\,, 
\end{equation}
where the parameter $\varepsilon$ characterizes the degree of deviation of this relation from the result of current algebra and PCAC ($\varepsilon=0$). It turns out that chiral corrections are significant for $\varepsilon$. In determining the quark masses, we used the FLAG average value $\varepsilon=0.79(6)$ $(N_f=2+1+1)$ \cite{FLAG:22}, which yields
\begin{equation}
\label{emkaon}
\tilde\Delta^2_{em}=2.21(8)\times 10^{-3}\,\mbox{GeV}^2.
\end{equation}

Explicit calculations performed above (and partly in Section\,7, where the value of the $\pi^0$ mass is obtained) allow us to express the constants $\Delta^2_{em}$ and $\tilde\Delta^2_{em}$ to next-to-leading order in $\delta$ through the parameters of the effective theory, namely
\begin{eqnarray}
    \Delta^2_{em}&=&2\frac{e^2}{F^2}C\left(1-2\hat m\Delta'\right) +2e^2\mu_{\pi^\pm}^2(k_1^r+4k_4^r) \nonumber \\
    &-&\frac{e^2}{16\pi^2} \mu_{\pi^\pm}^2\left(3\ln\frac{\mu_{\pi^\pm}^2}{\mu^2}-4\right),  \\
    \tilde\Delta^2_{em}&=&2\frac{e^2}{F^2}C\left[1-(m_s+\hat m)\Delta'\right] \nonumber \\
    &-&\frac{e^2}{16\pi^2} \hat\mu_{K}^2\left(3\ln\frac{\hat\mu_{K}^2}{\mu^2}-4\right)\nonumber \\
    &-&\frac{4}{3}e^2\left[\hat\mu_K^2(k_2^r-k_1^r-6k_4^r)-\mu_{\pi^\pm}^2k_3^r \right]. 
 \end{eqnarray}
The numerical values of these quantities coincide with (\ref{empion}) and (\ref{emkaon}). This coincidence is certainly not accidental. It reflects the internal self-consistency of the procedure for fixing the quark masses on the one hand and the subsequent determination of the electromagnetic parameters $k^r_1,\ldots,k^r_4$ on the other.


\section{Neutral mesons: kinetic term and decay constants }

When considering neutral states, difficulties associated with their mixing arise. There are several reasons for this. Mixing is caused by the effects of explicit violation of the $SU(2)$ and $SU(3)$ symmetries of strong interactions. Electromagnetic interactions also lead to mixing. In this section, we eliminate mixing in the kinetic part of the effective Lagrangian $\mathcal L=\mathcal L^{(0)}+\mathcal L^{(1)}$ by redefining the neutral fields. The diagonalization of its mass part is performed in the next section. We also derive expressions for the decay constants, which contain, in addition to chiral corrections, also an electromagnetic contribution.

In the most general case, the neutral field $\phi$ can be represented by its components taking values in the algebra of the $U(3)$ group or more specifically in the subset of the diagonal hermitian generators $\{\lambda_a\}_{a=0,8,3}$ or their linear combinations $\{\lambda_i\}_{i=u,d,s}$
\begin{equation}
\label{Phi-field}
\phi=\!\!\!\!\sum_{a=0,8,3}\!\!\phi_a\lambda_a=\!\!\!\!\sum_{i=u,d,s}\!\!\phi_i\lambda_i .
\end{equation}  
There is a simple relation between these two bases:
\begin{equation}
\lambda_i =\frac{1}{2} \sum_{a=0,8,3} O_{ia} \lambda_a ,
\end{equation}
where
\begin{equation}
\label{O}
 O=\frac{1}{\sqrt 3}\left( 
\begin{array}{ccc}
\sqrt 2 & 1 & \sqrt 3   \\
\sqrt 2 & 1 & -\sqrt 3 \\
\sqrt 2 & -2 & 0 \\
\end{array}
\right), \quad  O^{-1} =\frac{1}{2} O^T.
\end{equation}
The matrix $O$ has the property: $\sum_a O_{ia}O_{ja}=2\delta_{ij}$. It is also clear from (\ref{Phi-field}) that
\begin{equation}
\label{Fields}
\phi_i=\!\!\! \sum_{a=0,8,3}\!\!\! O_{ia}\phi_a, \quad \phi_a=\!\!\!\sum_{i=u,d,s}\!\!\! O^{-1}_{ai}\phi_i, 
\end{equation}
and $\langle\phi^2\rangle \!=\!2\sum\phi_a^2\!=\!\sum\phi_i^2$.

One can see that in the special case $\Lambda_1=0$, $e=0$, the kinetic part of the Lagrangian $\mathcal L$ takes the standard form if we rescale the flavor components:
\begin{equation}
\label{rescaling}
f_i\phi_i=\phi_i^R,
\end{equation} 
where $f_i=1+m_i\Delta'$.

The new dimensional field $\phi^R$ can also be characterized by its components in any of the $\lambda$-matrix bases
\begin{equation}
\label{Rfields}
\phi^R=\!\!\!\!\sum_{i=u,d,s}\!\!\phi_i^R\lambda_i=\!\!\!\!\sum_{a=0,8,3}\!\!\phi_a^R\lambda_a.
\end{equation}  

This is sufficient to establish a relationship between the fields $\phi_a$ and $\phi^R_a$, which in fact has a non-diagonal form. Thus, we need to find the symmetric matrix $f_{ab}$ and its inverse $f^{-1}_{ab}$ that  interpolates between these fields
\begin{equation}
\label{mix}
\phi_a^R=\sum_b f_{ab} \phi_b, \quad \phi_a=\sum_b (f^{-1})_{ab} \phi_b^R.
\end{equation} 

To do this, let us consider the sum for $\phi^R$:
\begin{equation}
\sum_i f_i \phi_i \lambda_i =\sum_{i,b} f_i O_{ib}\phi_b\lambda_i=\frac{1}{2}\sum_{i,b,a} f_i O_{ib} O_{ia}\phi_b\lambda_a.
\end{equation}
Comparing the result with the right-hand side of Eq.\,(\ref{Rfields}) we find that
\begin{equation}
\label{matrixF}
f_{ab}=\frac{1}{2}\sum_{i=u,d,s}\!\!\! O_{ia} O_{ib}  f_i.
\end{equation}
In components this gives
\begin{eqnarray}
\label{elementsF}
 &&f_{00}\!=\!\frac{1}{3}\left(f_u\!+\!f_d\!+\!f_s\right) 
\!=\!F\!\left[1\!+\!\frac{\Delta'}{3}(m_u\!+\!m_d\!+\!m_s)\right], \nonumber \\
 &&f_{08}\!=\!\frac{f_u\!+\!f_d\!-\!2f_s}{3\sqrt 2}
 \!=\!\frac{F\Delta'}{3\sqrt 2}\left(m_u\!+\!m_d\!-\!2m_s\right), \nonumber \\
&&f_{03}\!=\!\sqrt 2 f_{38}\!=\! \frac{1}{\sqrt 6}\left(f_u\!-\!f_d\right)\!=\!\frac{F\Delta'}{\sqrt 6}(m_u\!-\!m_d), \nonumber \\
&&f_{88}\!=\!\frac{1}{6}\left(f_u\!+\!f_d\!+\!4f_s \right) 
\!=\!F\!\left[1\!+\!\frac{\Delta'}{6}(m_u\!+\!m_d\!+\!4m_s)\right], \nonumber \\
&&f_{33}\!=\!\frac{1}{2}\left(f_u\!+\!f_d\right)\!=\! F(1\!+\!\hat m \Delta'). 
\end{eqnarray}

Starting from equation (\ref{Phi-field}) and acting in a similar way, we find
\begin{equation}
\sum_i \frac{\phi_i^R}{f_i} \lambda_i =\sum_{i,b} \lambda_i O_{ib}\frac{\phi^R_b}{f_i}=\frac{1}{2}\sum_{i,b,a} O_{ib} O_{ia} \frac{\phi_b^R}{f_i} \lambda_a.
\end{equation}
Therefore
\begin{equation}
\mathcal (f^{-1})_{ab}=\frac{1}{2}\sum_{i=u,d,s}\!\!\! O_{ia} O_{ib}  \frac{1}{f_i},
\end{equation}
i.e., the elements of the inverse matrix are obtained from (\ref{elementsF}) by replacing $f_i\to 1/f_i$.    

As a result, Eq.\,(\ref{mix}) gives \cite{Osipov:23}
\begin{eqnarray}
\label{mixphi}
\phi_0 &\!=\!& \frac{\phi_0^R}{f_0}\!+\! \left(\frac{1}{f_u}\!-\!\frac{1}{f_d} \right) \frac{\phi_3^R}{\sqrt 6}  
\!+\! \left(\frac{1}{f_u}\!+\!\frac{1}{f_d}\!-\!\frac{2}{f_s}\right) \frac{\phi_8^R}{3\sqrt 2},  \nonumber \\            
\phi_8 &\!=\!& \frac{\phi_8^R}{f_8} \!+\! \left(\frac{1}{f_u}\!-\!\frac{1}{f_d} \right) \frac{\phi_3^R}{2\sqrt 3} 
\!+\!\left(\frac{1}{f_u}\!+\!\frac{1}{f_d}\!-\!\frac{2}{f_s}\right) \frac{\phi_0^R}{3\sqrt 2}, \nonumber\\
\phi_3 &\!=\!& \frac{\phi_3^R}{f_3} \!+\! \left(\frac{1}{f_u}\!-\!\frac{1}{f_d}\right)
\frac{\phi_8^R \!+\!\sqrt 2 \phi_0^R}{2\sqrt 3},               
\end{eqnarray}
where 
\begin{eqnarray}
\label{f083}
f_0^{-1}&=&\frac{1}{3} \left(f_u^{-1} \!+\!f_d^{-1}\! +\!f_s^{-1}\right), \nonumber \\
f_8^{-1}&=&\frac{1}{6} \left(f_u^{-1} \!+\!f_d^{-1}\! +\!4f_s^{-1}\right),  \nonumber \\
f_3^{-1}&=&\frac{1}{2} \left(f_u^{-1}\!+\!f_d^{-1} \right). 
\end{eqnarray}
A distinctive feature of the transition from fields $\phi_a$ to $\phi^R_a$ is the emergence of mixing. This is due to the explicit violation of both isospin and flavor $SU(3)$ symmetries. 

Now the entire kinetic part of the Lagrangian $\mathcal L$ can be rewritten (in the NLO approximation) in terms of the fields $\phi^R_a$ as follows 
\begin{equation}
\label{nkin}
\mathcal L_{kin}^{(n)}=\frac{1}{2}\!\sum_{a,b=0,8,3}\!\!\partial\phi^R_a (1+A)_{ab}\partial\phi^R_b ,
\end{equation}
where the symmetric matrix $A$ is given by
\begin{equation}
\label{A}
A=\left(\!\!
\begin{array}{ccc}
\Lambda_1 &\! 0 &\! 0    \\
\! 0 &\!\! 0 &\! 0 \\
\! 0 &\!\! 0 &\! 0 \\
\end{array}
\!\!\right)\!+\!
\frac{2}{9}e^2(k_1^r+2k_2^r)\!\left(\!\!\! 
\begin{array}{ccc}
4 &\!\! \sqrt{2} &\!\! \sqrt{6} \\
\sqrt{2} &\!\! 3 &\!\! \sqrt{3} \\
\sqrt{6} &\!\! \sqrt{3} &\!\! 5 \\
\end{array}
\!\!\right).
\end{equation}
The linear transformation $\phi^R=(1-A/2)\tilde \phi^R$ brings $(\ref{nkin})$ to canonical form, which finally determines the relation between the field variables at NLO 
\begin{equation}
\phi_a=\sum_{b=0,8,3} \tilde f^{-1}_{ab} \tilde \phi_b^R,
\end{equation}
where
\begin{eqnarray}
&&\tilde f^{-1}_{00}\!=\!\frac{1}{F}\!\left[1\!-\!\frac{\Lambda_1}{2}\!-\!\frac{\Delta'}{3}(m_u\!+\!m_d\!+\!m_s)   
\!-\!\frac{4e^2}{9}(k_1^r\!+\!2k_2^r)\right], \nonumber \\
&&\tilde f^{-1}_{08}\!=\!\frac{1}{3\sqrt 2 F}\!\left[\Delta'(2m_s\!-\!m_d\!-\!m_u)\!-\!\frac{2e^2}{3}(k_1^r\!+\!2k_2^r) \right], \nonumber \\
&&\tilde f^{-1}_{03}\!=\!\sqrt 2\tilde f^{-1}_{38}\!=\!\frac{1}{\sqrt 6 F}\!\left[\Delta'(\!m_d\!-\!m_u)\!-\!\frac{2e^2}{3}(k_1^r\!+\!2k_2^r) \right], \nonumber \\
&&\tilde f^{-1}_{88}\!=\!\frac{1}{F}\!\left[1\!-\!\frac{\Delta'}{6}(m_u\!+\!m_d\!+\!4m_s)   
\!-\!\frac{e^2}{3}(k_1^r\!+\!2k_2^r)\right], \nonumber \\
&&\tilde f^{-1}_{33}\!=\!\frac{1}{F}\!\left[1\!-\!\hat m \Delta'\!-\!\frac{5e^2}{9}(k_1^r\!+\!2k_2^r) \right].
\end{eqnarray}

To summarize the calculations performed above, we present expressions for the decay constants. They differ from similar expressions $F_{ab}$ obtained in \cite{Goity:02} (see Eq.\,(22) there) only in that they additionally include the contribution of electromagnetic interactions
\begin{eqnarray}
\tilde f_{00}&=&F_{00}+\frac{4}{9}F e^2(k_1^r+2k_2^r), \nonumber \\
\tilde f_{08}&=&F_{08}+\frac{\sqrt 2}{9}F e^2(k_1^r+2k_2^r), \nonumber \\
\tilde f_{03}&=&\sqrt 2 \tilde f_{38}=F_{03}+\frac{2}{3\sqrt 6}F e^2(k_1^r+2k_2^r), \nonumber \\
\tilde f_{88}&=&F_{88}+\frac{1}{3}F e^2(k_1^r+2k_2^r), \nonumber \\
\tilde f_{33}&=&F_{33}+\frac{5}{9}F e^2(k_1^r+2k_2^r).
\end{eqnarray}


\section{Weak-decay coupling constants in the singlet-octet bases}
\label{s6}

The meson decay constants $F^a_P$ are determined through the matrix elements of the axial-vector currents $A_\mu^a=\bar q\gamma^\mu\gamma_5\frac{\lambda^a}{2}q$
\begin{equation}
\langle 0| A_\mu^a(0) |P(p)\rangle =ip_\mu F^a_P.
\end{equation}
To find $F^a_P$, it is necessary to establish a connection between the physical states $P=\eta',\eta',\pi^0$ and the bare fields $\phi_a$: $P=\sum_a F^a_P\phi_a$. The physical states $P$ are eigenvectors of the mass matrix, and to find them, an additional transformation of the fields $\tilde\phi^R_a$ is required $P=R^{-1}\tilde\phi^R$, which is carried out by the orthogonal matrix $R$, diagonalizing the quadratic form of the mass part of the Lagrangian. In this section, we obtain the decay constants $F^a_P$ at NLO, assuming that the rotation matrix $R$ is given by 
\begin{equation}
\label{R}
R=\left( 
\begin{array}{ccc}
\! \cos\theta & -\sin\theta & \epsilon'\cos\theta\!-\!\epsilon \sin\theta\!    \\
\! \sin\theta  & \cos\theta & \epsilon' \sin\theta\!+\!\epsilon \cos\theta\! \\
\! -\epsilon' & -\epsilon & 1 \!\\
\end{array}
\right).
\end{equation}
The matrix $R$ is parameterized by the three angles $\theta,\epsilon,\epsilon'$  \cite{Leutwyler:96b}. The angle $\theta$ is non-zero due to the violation of the unitary $SU(3)$ symmetry $(m_s-\hat m)\neq 0$. The mixing angles $\epsilon$, $\epsilon'$ have the first order in the violation of isospin symmetry, i.e., they contain the terms $\sim (m_d-m_u)$ and $\sim e^2$. We neglect higher order contributions in isospin symmetry breaking, i.e., $\epsilon^2$, $\epsilon'^2$, $\epsilon\epsilon'$, etc. 

Thus we find
\begin{equation}
F^a_P=(R^{-1}_{ab}\tilde f)^a_P,
\end{equation}
or  in components (with the accuracy adopted here)
\begin{eqnarray}
\label{constFaP}
&&F^0_{\eta'}\!=\!\tilde f_{00} \cos\theta\! +\! \tilde f_{08}\sin\theta, \nonumber \\
&&F^8_{\eta'}\!=\!\tilde f_{08} \cos\theta\! +\! \tilde f_{88}\sin\theta, \nonumber \\
&&F^0_{\eta}\!=\!-\tilde f_{00} \sin\theta\! +\! \tilde f_{08}\cos\theta, \nonumber \\
&&F^8_{\eta}\!=\!-\tilde f_{08} \sin\theta\! +\! \tilde f_{88}\cos\theta, \nonumber  \\
&&F^3_{\eta'}\!=\!\tilde f_{03} \cos\theta\! +\! \tilde f_{38}\sin\theta\! -\!\epsilon' \tilde f_{33},  \nonumber \\
&&F^3_{\eta}\!=\!-\tilde f_{03} \sin\theta\! +\! \tilde f_{38}\cos\theta \!-\! \epsilon \tilde f_{33}, \nonumber \\
&&F^0_{\pi^0}\!=\!\tilde f_{03}\!+\!\tilde f_{00}(\epsilon'\!\cos\theta\!-\!\epsilon\sin\theta )\!+\! \tilde f_{08}(\epsilon'\! \sin\theta\!+\!\epsilon\cos\theta ), \nonumber \\
&&F^8_{\pi^0}\!=\!\tilde f_{38}\!+\!\tilde f_{08}(\epsilon'\!\cos\theta\!-\!\epsilon\sin\theta )\!+\! \tilde f_{88}(\epsilon'\! \sin\theta\!+\!\epsilon\cos\theta ), \nonumber \\
&&F^3_{\pi^0}\!=\!\tilde f_{33}=f_{\pi^0}. 
\end{eqnarray}

In (\ref{constFaP}), it is necessary to take into account only the terms that do not exceed the accuracy of our calculations. 
This also applies to the mixing angles, which should be represented as the sum of the LO contribution and the NLO correction to it: $\theta=\theta_0+\delta\theta$, $\epsilon =\epsilon_0+\delta\epsilon$, $\epsilon' =\epsilon_0'+\delta\epsilon'$. 

To lowest order in $\delta$, we have $\tilde f_{08}=\tilde f_{03}=\tilde f_{38}=0$ and then for $\eta$-$\eta'$ system we arrive to the standard pattern with one mixing angle $\theta_0$. In contrast, taking into account the NLO correction leads to a scheme with two mixing angles for $\eta$-$\eta'$ system. To verify this, we expand (\ref{constFaP}) into a series in $\delta$. Up to NLO terms, we find:
\begin{eqnarray}
\label{FaPsol}
&&F^0_{\eta'}=\tilde f_{00}\cos\theta_0-F\Delta\theta_{-}\sin\theta_0,   \nonumber\\
&&F^0_{\eta}=-\tilde f_{00}\sin\theta_0 -F\Delta\theta_{-} \cos\theta_0,  \nonumber\\ 
&&F^8_{\eta'}=\tilde f_{88}\sin\theta_0 +F\Delta\theta_{+}\cos\theta_0, \nonumber \\
&&F^8_{\eta}=\tilde f_{88}\cos\theta_0 -F\Delta\theta_{+}\sin\theta_0, 
\end{eqnarray}
where
\begin{equation}
\Delta\theta_{\pm}\equiv\delta\theta \pm \frac{\tilde f_{08}}{F}.
\end{equation}

The $\delta$-corrections to the leading order result allows one to distinguish two mixing angles $\vartheta_8$ and $\vartheta_0$, which are often used in the phenomenological analysis of $\eta$-$\eta'$ data \cite{Feldmann:98,Escribano:05,Escribano:16}. Indeed, from  Eq.\,(\ref{FaPsol})  it follows
\begin{eqnarray}
\label{angle8}
F^8_{\eta'}&\!=\!&\tilde f_{88} \left( \sin\theta_0\! +\!\frac{F}{\tilde f_{88}}\Delta\theta_+\cos\theta_0\right) \nonumber \\
&\!\simeq\!&\tilde f_{88}\sin (\theta_0\!+\!F\Delta\theta_+/\tilde f_{88})\! =\!  \tilde f_{88} \sin\vartheta_8,  \\
\label{angle0}
F^0_{\eta}&\!=\!& -\!  \tilde f_{00} \left( \sin\theta_0\! +\!\frac{F}{\tilde f_{00}}\Delta\theta_-\cos\theta_0\right)
\nonumber \\
&\!\simeq\!&-\tilde f_{00}\sin (\theta_0\!+\!F\Delta\theta_-/\tilde f_{00}) \!=\! -\tilde f_{00}\sin\vartheta_0. 
\end{eqnarray}
That gives 
\begin{eqnarray}
\vartheta_8&=&\theta_0+F\Delta\theta_+/\tilde f_{88}, \nonumber \\
\vartheta_0&=&\theta_0+F\Delta\theta_-/\tilde f_{00}.
\end{eqnarray}
Numerical estimates of $\vartheta_8$ and $\vartheta_0$ will be made below when we fix the angles of matrix $R$. What we can do now is to estimate the values of the decay constants: 
\begin{eqnarray}
\label{f0088}
\tilde f_{88}&=&1.28 f_{\pi}=118.2\,\mbox{MeV}, \nonumber \\
\tilde f_{00}&=&1.07 f_{\pi}=99.1\,\mbox{MeV}.
\end{eqnarray}
The first relation is in perfect agreement with estimates $F_8=1.28 f_{\pi}$ \cite{Leutwyler:98} and  $F_8=1.27(2) f_{\pi}$ \cite{Escribano:16}. The second relation cannot be verified phenomenologically, since the constant $\tilde f_{00}$ depends on the running scale $\mu_{\mbox{\tiny QCD}}$. Our estimate is made at $\mu_{\mbox{\tiny QCD}}=2\,\mbox{GeV}$. 

One should also note the agreement between the values of (\ref{f0088}) and recent estimates of these low-energy constants in \cite{Oller:15}, where the NLO fit yields $F_0=99.7(3.6)(1.6)\,\mbox{MeV}$ and $F_8=113.5(0.3)(4.2)\,\mbox{MeV}$ for NLOFit-B. The latter estimates were obtained on the basis of alternative considerations. While we proceeded from the phenomenological values of the $\pi^+$, $K^+$, $K^0$ meson masses and the decay constants $f_\pi$, $f_K$, in \cite{Oller:15} the masses of the $\eta$, $\eta'$ and $K$ mesons were used for this purpose and the values of $f_\pi$ and $f_K$ were not involved, since they are poorly reproduced at next-to-leading order in $\delta$ expansion in the absence of the electromagnetic interaction. Taking into account the electromagnetic interaction eliminates the problem of describing these constants at NLO, and the coincidence of the results can be considered as an indicator of the internal consistency of the emerging picture.


\section{Neutral mesons: masses and mixing angles}

Before moving on to the diagonalization of the mass part of the Lagrangian, we note that in the new variables $\tilde\phi_R$ the mass matrix additionally receives a contribution from the anticommutator\begin{eqnarray}
    \phi^T_R M^2\phi_R&=&\tilde \phi^T_R \left(M^2-\frac{1}{2}\left\{A,\mu^2 \right\}\right)\tilde \phi_R  \nonumber \\
    &\equiv& \tilde\phi^T_R \tilde M^2\tilde \phi_R.
\end{eqnarray}
As a result, the gluon anomaly contained in the singlet component $\mu^2_{00}$ affects the contribution of electromagnetic interactions in the NLO approximation.

To find the masses of $\pi^0$, $\eta$ and $\eta'$ mesons, it is necessary to diagonalize the quadratic form
\begin{equation}
    L_m=-\frac{1}{2}\sum_{a=0,3,8}\tilde\phi^R_{a} \tilde M^2_{ab}\tilde\phi^R_{b}.
\end{equation}

The symmetric matrix $\tilde M_{ab}^2$ includes the leading order contribution $\mu^2_{ab}$, as well as the NLO correction $\delta\mu^2_{ab}$
\begin{equation}
\label{mass2}
\tilde M_{ab}^2=\mu^2_{ab}+\Delta\mu^2_{ab}+\Delta\tilde\mu^2_{ab}\equiv\mu^2_{ab}+\delta\mu^2_{ab}.
\end{equation}

The part $\Delta\mu^2_{ab}$ collects corrections induced by strong interactions, and the last term $\Delta\tilde\mu^2_{ab}$ by electromagnetic ones. The contributions $\mu^2_{ab}$ and $\Delta\mu^2_{ab}$ are well known (see, for example, \cite{Goity:02,Osipov:23}). Nevertheless, we present them here for completeness:
\begin{eqnarray}
    \mu_{00}^2&=&\frac{2}{3}B_0(2\hat m+m_s)+M^2_0, \nonumber \\
    \mu_{08}^2&=&\frac{\sqrt{8}}{3}B_0(\hat m-m_s), \nonumber \\
    \mu_{03}^2&=&\sqrt{\frac{2}{3}}B_0(m_u-m_d), \nonumber  \\
    \mu_{88}^2&=&\frac{2}{3}B_0(\hat m+2m_s), \nonumber \\
    \mu_{38}^2&=&\frac{1}{\sqrt{3}}B_0(m_u-m_d), \nonumber \\
    \mu_{33}^2&=&2B_0\hat m.
\end{eqnarray}
Note that the constant $M^2_0=\mathcal{O}(\delta)$ is proportional to the topological susceptibility of the pure gluon theory and represents the contribution of the $U(1)_A$ gluon anomaly.
 
The elements of the matrix $\Delta\mu^2_{ab}$ have the form
\begin{eqnarray}
    \Delta\mu_{00}^2&=&\frac{2}{3}B_0\left[2(2\hat m^2+m_s^2)\Delta+(2\hat m +m_s)\rho \right]-M_0^2\Lambda_1, \nonumber \\
    \Delta\mu_{08}^2&=&\frac{\sqrt{2}}{3}B_0 (\hat m-m_s)\left[4(\hat m+m_s)\Delta+\rho\right],\nonumber \\
    \Delta\mu_{03}^2&=&\sqrt{\frac{2}{3}}B_0 (m_u-m_d)\left(4\hat m\Delta+\frac{\rho}{2}\right), \nonumber \\
    \Delta\mu_{88}^2&=&\frac{4}{3}B_0\left(\hat m^2+2m_s^2\right)\Delta,\nonumber \\
    \Delta\mu_{38}^2&=&\frac{2}{\sqrt{3}}B_0(m_u^2-m_d^2)\Delta, \nonumber \\
    \Delta\mu_{33}^2&=&4B_0\hat m^2\Delta,
\end{eqnarray}
where $\rho=2\Lambda_2-\Lambda_1-M_0^2\Delta'/B_0$. The parameters $\Lambda_1$ and $\Lambda_2$ characterize the magnitudes of the contributions associated with the violation of Zweig's rule.

The electromagnetic contribution $\Delta\tilde\mu^2_{ab}$ was calculated in \cite{Prades:97}, where, however, the authors neglected the $U(1)_A$ gluon anomaly. As will be shown below, it is this correction ($\propto M_0^2 $) that plays the most significant role in $\Delta\tilde\mu^2_{ab}$, leading to a restoration of isospin symmetry  at NLO:
\begin{eqnarray}
\label{emM2}
  \Delta\tilde\mu^2_{00}&=&-\frac{8}{9}e^2\left[ M_0^2(k_1^r+2k^r_2)+\frac{1}{3}(5\hat m+m_s)\tilde B_0 \right],  \nonumber \\
  \Delta\tilde\mu^2_{08}&=&-\frac{\sqrt{2}}{9}e^2\left[ M_0^2(k_1^r+2k^r_2)+\frac{4}{3}(5\hat m-2m_s) \tilde B_0 \right],  \nonumber \\
 \Delta\tilde\mu^2_{03}&=&-\frac{\sqrt{6}}{9}e^2\left[ M_0^2(k_1^r+2k^r_2)+4\hat m \tilde B_0 \right], \nonumber \\
  \Delta\tilde\mu^2_{88}&=&
  -e^2\frac{4}{27}(5\hat m +4m_s)\tilde B_0,   \nonumber \\
  \Delta\tilde\mu^2_{38}&=&
  -\frac{4}{3\sqrt{3}}e^2 \hat m\tilde B_0,   \nonumber \\
  \Delta\tilde\mu^2_{33}&=&
  -\frac{20}{9}e^2 \hat m \tilde B_0,
\end{eqnarray}
where for brevity we put $\tilde B_0\equiv B_0 (k_1^r+2k^r_2-2k_3^r)$. 

Since we neglect higher order terms in isospin symmetry breaking, diagonalization does not affect the value of the neutral pion mass
\begin{equation}
  m^2_{\pi^0}=\tilde M^2_{33}=\mu^2_{\pi^\pm}\left(1+2\hat m\Delta\right)-\frac{20}{9}e^2 \hat m  \tilde B_0.  
\end{equation}
Numerically this gives $m_{\pi^0}=135\,\mbox{MeV}$.  

As a result, to fix the three unknown parameters $M_0$, $\Lambda_1$ and $\rho$, only two phenomenological values remain -- these are the masses of $\eta$ and $\eta'$ mesons. Considering that the fitting of $\Lambda_1$ (in the absence of electromagnetic interactions) yields $\Lambda_1=-0.04\pm 0.06\pm 0.13$ \cite{Oller:15}, we will study three alternative options: $\Lambda_1=0,\pm 0.15$. Then, using the expressions obtained by diagonalizing the mass matrix for the squares of the masses of $\eta$ and $\eta'$ mesons in the NLO approximation
\begin{equation}
\label{meta}
   m_{\eta,\eta'}^2=\mu^2_{\eta,\eta'}+\delta\mu^2_{\eta,\eta'},   
\end{equation}
where
\begin{eqnarray}
  \mu^2_{\eta,\eta'}&=&\frac{1}{2}\left(\mu^2_{00}+\mu^2_{88}\mp\sqrt{(\mu^2_{00}-\mu^2_{88})^2+(2\mu^2_{08})^2}\right)\nonumber \\
  \delta\mu^2_{\eta,\eta'}&=&\frac{1}{2}\left[\delta\mu^2_{00}+\delta\mu^2_{88}\right. \\
  &\mp&\left. \frac{(\delta\mu^2_{00}-\delta\mu^2_{88})(\mu^2_{00}-\mu^2_{88})+4\delta\mu_{08}^2\mu^2_{08}}{\sqrt{(\mu^2_{00}-\mu^2_{88})^2+(2\mu^2_{08})^2}}  \right] \nonumber , 
\end{eqnarray}        
one can fix the two remaining parameters. The result is presented in Table\,I.

\begin{table*}
\label{tab1}
\caption{The mixing angles $\theta$, $\epsilon$, $\epsilon'$, and the parameters $M_0$, $\Lambda_2$, $\rho$ obtained by fitting the squares of the meson masses $m^2_\eta$ and $m^2_{\eta'}$ to their phenomenological values under the assumption that $\Lambda_1=0,\pm 0.15$ (these input data are marked $(^*)$). The error bars indicated in (a), (b), and (c) are mainly determined by the accuracy with which the quark masses and $B_0$ are known.} 
\begin{footnotesize}
\begin{tabular*}{\textwidth}{@{\extracolsep{\fill}}cccccccc@{}}
\hline
\hline 
\multicolumn{1}{l}{\ \ \ \ Fit}
& \multicolumn{1}{c}{$\Lambda_1$}
& \multicolumn{1}{c}{$\Lambda_2$}
& \multicolumn{1}{c}{$M_0\,[\mbox{MeV}]$}
& \multicolumn{1}{c}{$\rho$}
& \multicolumn{1}{c}{$\theta$}
& \multicolumn{1}{c}{$\epsilon$}
& \multicolumn{1}{c}{$\epsilon'$}
\\
\hline
\\[-9pt] 
(a)
& $0.15^*$ 
& $0.410(5)$
& $1044(3)$
& $-1.065(10)$
& $-9.48(13)^\circ$
& $0.60(2)^\circ$
& $-0.058(5)^\circ$ 
\\
(b)
& $0^*$
& $0.248(5)$
& $954(2)$
& $-0.950(6)$
& $-10.39(12)^\circ$
& $0.58(2)^\circ$
& $-0.037(5)^\circ$ 
\\
(c)
& $-0.15^*$
& $0.128(2)$
& $879(1)$
& $-0.824(3)$
& $-11.04(11)^\circ$
& $0.55(2)^\circ$
& $-0.013(6)^\circ$
\\
\cite{Goity:02} NLO No.1
& $0.19(1)$
& $0.74(2)$
& $1047(5)$
& $-0.950(10)$
& $-10.6^\circ$
& $0.85(5)^\circ$
& $\simeq 0.3^\circ$
\\
\cite{Oller:15} NLOFit-C
& $-0.06(4)(2)$
& $0.17(19)(25)$
& $835.7$
& $-$
& $-10.6(2.4)(3.3)^\circ$
& $-$
& $-$
\\
\hline
\hline
\end{tabular*}
\end{footnotesize} 
\end{table*}

For comparison, Table\,I includes the estimates given in \cite{Goity:02} for the NLO No.1 case and in \cite{Oller:15} for the NLOFit-C case. They are obtained under assumptions similar to those used here.

Electromagnetic interactions, as expected, have virtually no effect on the magnitude of the $\eta$-$\eta'$ mixing angle $\theta$. Here, strong interactions play a dominant role. In leading order, its value $\theta_0$ is determined by the expression
\begin{equation}
    \tan 2\theta_0 =\frac{\sqrt 8}{1+\sqrt{2}M_0^2/\mu^2_{08}}
\end{equation}
from where for three different values of $\Lambda_1$ we find
\begin{eqnarray}
    \theta_0&=&-12.98(1)^\circ,\quad \delta\theta=3.50(14)^\circ , \quad \mbox{(a)} \nonumber \\
    \theta_0&=&-15.59(2)^\circ,\quad \delta\theta=5.20(14)^\circ , \quad \mbox{(b)}  \\
    \theta_0&=&-18.25(3)^\circ,\quad \delta\theta=7.21(14)^\circ . \quad\,\, \mbox{(c)} \nonumber
\end{eqnarray}
Here we also present the contribution of the NLO correction $\delta\theta$
\begin{equation}
    \delta\theta=\frac{\mu^2_{08}(\delta\mu_{88}^2-\delta\mu_{00}^2)+\delta\mu^2_{08}(\mu_{00}^2-\mu_{88}^2)}{(\mu_{00}^2-\mu_{88}^2)^2+(2\mu^2_{08})^2},
\end{equation}
in which electromagnetic interactions account for only $\delta\theta_{\mbox{\footnotesize{em}}}\simeq 0.34(2)^\circ$. This is about 3\% of the full value of the angle $\theta$. Note also that the values of $\theta$ given in Table\,I are consistent with estimates \cite{Oller:15,Goity:02,Kroll:05}.
 
 For a scheme with two mixing angles, we find from (\ref{angle8}) and (\ref{angle0}) that
\begin{equation}
\label{08two}
\vartheta_0= -1.7^\circ, \quad \vartheta_8 = -21^\circ.
\end{equation} 
The given numerical values correspond to the case (c) of Table\,I, which we consider to be the most preferable. Since the angle $\vartheta_0$ is small, the $\eta$ meson is nearly pure octet state. This can be seen in more detail by calculating the projections of the $\eta$ state onto the singlet-octet basis: $F_\eta^0=0.04\, F$, $F_\eta^8=1.18\, F$ and $F^3_\eta=-0.02\, F$.

The result (\ref{08two}) is consistent with the data obtained from the phenomenological analysis of various electromagnetic decays performed in \cite{Escribano:05} using the two-angle mixing description of the $\eta$-$\eta'$ system and including the connection with the gluon sector via anomalies, where the authors arrive at $\vartheta_0= -2.4(1.9)^\circ$, $\vartheta_8=-23.8(1.4)^\circ$.



Let us now turn to a discussion of the results obtained for the mixing angles $\pi^0$-$\eta$ and $\pi^0$-$\eta'$. These angles differ from zero only due to isospin symmetry breaking. Their magnitude is strongly influenced by the $U(1)$  anomaly of QCD: The pseudoscalar spectrum would not show signs of isospin symmetry if the axial Ward identities did not contain an anomalous term \cite{Gross:79}. Indeed, the ratio of quark masses $m_u/m_d=0.455(8)$ strongly violates the $SU(2)$ symmetry and only due to the $U(1)$ anomaly the angle of $\pi^0$-$\eta$ mixing turns out to be small
\begin{equation}
\label{barep0}
\bar\epsilon_0=\frac{\sqrt{3}}{4}\frac{m_d-m_u}{m_s-\hat m}=0.012.
\end{equation}

In the leading order of the $\delta$-expansion, due to $\eta$-$\eta'$ mixing, the angles $\epsilon_0$ and $\epsilon_0'$ acquire an additional factor depending on $\theta_0$ \cite{Leutwyler:96b}   
\begin{eqnarray}
\label{ep0}
\epsilon_0&=&\bar\epsilon_0\cos\theta_0\frac{\cos\theta_0-\sqrt{2}\sin\theta_0}{\cos\theta_0+\sin\theta_0/\sqrt{2}}, \\
\label{epp0}
\epsilon_0'&=&\bar\epsilon_0\sin\theta_0\frac{\sin\theta_0+\sqrt{2}\cos\theta_0}{\sin\theta_0-\cos\theta_0/\sqrt{2}}.
\end{eqnarray}

It follows that for $\theta_0=0$: $\epsilon_0=\bar\epsilon_0$, and $\epsilon_0' =0$. For the finite values of $\theta_0$ found above, the angle $\epsilon_0'$ becomes nonzero $\epsilon_0'\simeq 0.3\, \bar\epsilon_0$, and the angle $\epsilon_0$ increases to values $\epsilon_0=\bar\epsilon_0\times [1.54\, \mbox{(a)},$ $1.67\,\mbox{(b)},\ 1.82\,\mbox{(c)}]$, making the theoretical estimate of the decay width $\eta\to 3\pi$ unacceptably large \cite{Leutwyler:96b}. The leading approximation also poorly describes the mass of the $\eta$-meson \cite{Georgi:94}, whose value turns out to be 9\% smaller than the phenomenological one. To avoid these difficulties, one should turn to the consideration of NLO corrections, including those of electromagnetic origin.

Indeed, if we neglect the electromagnetic contributions (\ref{emM2}), then for the angles $\epsilon$ and $\epsilon'$ we obtain a result close to \cite{Goity:02}: $\epsilon =0.82(2)^\circ$, $\epsilon'= 0.145(5)^\circ$ (a). Taking into account the electromagnetic corrections, as can be seen from Table\,I, introduces a significant adjustment to this estimate. The reason for the enhancement of the electromagnetic contribution is the $U(1)_A$ anomaly, which is present in the elements of the mass matrix $\Delta\tilde \mu_{00}^2$, $\Delta\tilde\mu_{08}^2$ and $\Delta\tilde\mu_{03}^2$. And if in the first two its role is less noticeable, then in $\Delta\tilde\mu_{03}^2$ it dominates and becomes important when calculating the value of the NLO correction to the angles $\epsilon$ and $\epsilon'$.

As a result, the electromagnetic part of the NLO correction $\delta\epsilon_{\mbox{\footnotesize em}}$ to the $\pi^0$-$\eta$ mixing angle $\epsilon$ compensates for the enhancement caused by $\eta$-$\eta'$ mixing in (\ref{ep0}), and thus returns us to the result of Gross, Treiman and Wilczek $\epsilon\simeq0.56^\circ$. The same thing happens with the value of the $\pi^0$-$\eta'$ mixing angle $\epsilon'$, which, as can be seen from Table\,I, is close to zero. It can be said that the axial anomaly enhances the isospin-destroying electromagnetic forces, but only to the extent that it suppresses the effects of singlet-octet mixing in angles $\epsilon$ and $\epsilon'$.

To see this, we write out an explicit expression for the angle $\epsilon$ in the NLO approximation
\begin{eqnarray}
\label{eps}
\epsilon&=&\epsilon_0\left[1-\delta\theta\,\frac{\sin\theta_0+\sqrt{2}\cos\theta_0}{\cos\theta_0-\sqrt{2}\sin\theta_0}-\frac{\delta\mu^2_\eta-\delta\mu^2_{33}}{\mu^2_\eta-\mu_{33}^2}\right] \nonumber \\
&+&\frac{\delta\mu^2_{03}\sin\theta_0-\delta\mu^2_{38}\cos\theta_0}{\mu^2_\eta-\mu_{33}^2}.
\end{eqnarray}
Numerically $\epsilon\simeq 0.014-0.004=0.01=0.57^\circ$. Here the subtrahend is the electromagnetic contribution of the last term in (\ref{eps}), in which everything is determined by the term $\propto\Delta\tilde\mu^2_{03}$, i.e. the gluon anomaly. Thus, the NLO correction due to electromagnetic interactions enhanced by the gluon anomaly is about 20\% of the contribution of the leading approximation. About the same amount (with the same sign) is accounted for by the correction related to strong interactions, which ultimately gives $\epsilon\simeq \bar\epsilon_0$, but for $\theta\neq 0$.

Let us now turn to the analysis of $\pi^0$-$\eta'$ mixing, characterized by the angle $\epsilon'$. In the LO,  $\epsilon_0'\simeq 0.2^\circ\neq 0$. Taking into account the NLO correction, we obtain
\begin{eqnarray}
\label{epsp}
\epsilon'&=&\epsilon_0'\left[1+\delta\theta\,\frac{\cos\theta_0-\sqrt{2}\sin\theta_0}{\sin\theta_0+\sqrt{2}\cos\theta_0}-\frac{\delta\mu^2_{\eta'}-\delta\mu^2_{33}}{\mu^2_{\eta'}-\mu_{33}^2}\right] \nonumber \\
&-&\frac{\delta\mu^2_{03}\cos\theta_0+\delta\mu^2_{38}\sin\theta_0}{\mu^2_{\eta'}-\mu_{33}^2}.
\end{eqnarray}
The contribution of electromagnetic interactions $\delta\epsilon'_{\mbox{\footnotesize em}}$ not only dominates in $\propto\Delta\tilde\mu^2_{03}$, but also generally determines the magnitude of the correction $\delta\epsilon'$, which leads to an almost complete restoration of isospin symmetry:
\begin{eqnarray}
 \epsilon_0'+ \delta\epsilon'_{\mbox{\footnotesize em}}&\simeq& \epsilon_0'\left[1+\frac{e^2M_0^2(k_1^r+2k_2^r)}{3B_0(m_d-m_u)(1+\frac{1}{\sqrt{2}}\tan\theta_0)}\right]\nonumber \\
 &=&\epsilon_0'(1+\gamma'),
\end{eqnarray}
where $\gamma'=-1.022(30)$ (a) and $\gamma'=-0.790(25)$ (c). The reduction of the leading contribution with the $1/N_c$ correction would seem to indicate the existing difficulties with the $\delta$-expansion. However, it should be taken into account that the LO result $\epsilon_0'\simeq 0.3\bar\epsilon_0$ is almost three times smaller than the scale of isospin symmetry breaking $\bar\epsilon_0$. At $N_c=3$ this makes the contributions comparable. The small negative value of $\epsilon'$ given in the Table\,I is the result of taking into account the NLO correction from strong interactions.


\section{Conclusions}

We calculated the spectrum, mixing angles, and decay constants of the pseudoscalar meson nonet. The effective Lagrangian used collects all terms of the leading and next-to-leading order in the large $N_c$ expansion that are relevant to the problem studied. Particular attention was paid to the contribution of virtual photons to the self-energy of mesons and the related effects of isospin symmetry breaking. 

The main effect generated by the electromagnetic corrections to the meson self-energies concerns $\pi^0$-$\eta$ and $\pi^0$-$\eta'$ mixing:  Electromagnetic interactions lead to singlet-octet mixing. As a consequence, the QCD anomaly penetrates the self-energy of neutral states and enhances the impact of electromagnetic interactions on the angles $\epsilon$ and $\epsilon'$. As a result, the effects of strong ($m_u\neq m_d$) and electromagnetic violations of isospin symmetry either double each other (in case of $\epsilon$) or largely cancel each other out (in case of $\epsilon'$). It is interesting that in both cases the numerical result turns out to be very close to the values obtained in \cite{Gross:79}, with the only exception being that the $\eta$-$\eta'$ mixing angle $\theta$ in our consideration is different from zero, taking value of about $-10^\circ$, or $\vartheta_0\simeq -2^\circ$ and $\vartheta_8\simeq -21^\circ$ in the two mixing angle scheme. Thus, the role of the NLO corrections is reduced to almost complete compensation of the effects of additional isospin symmetry breaking associated with $\theta\neq 0$. This can be considered a consequence of the fact that the topological susceptibility of gluodynamics $\propto M_0^2$ turns out to be quite large (Recall that the result of \cite{Gross:79} arises in the limit $M_0\to\infty$). 

To summarize, the new manifestation of the QCD anomaly investigated in this paper not only successfully copes with the problem of describing the $\eta\to 3\pi$ decay, but also allows us to take a broader look at the result of the work \cite{Gross:79}. The latter concerns their conclusion that in QCD the largest source of nonelectromagnetic isospin violation that can be expected arises from the $\pi^0$-$\eta$ mixing, that would otherwise be unmixed. As follows from our study, this conclusion is valid even in the presence of electromagnetic interactions and $\eta$-$\eta'$ mixing. Moreover, we have shown that the anomaly, in NLO approximation, explicitly participates in the restoring of isospin symmetry by enhancing the electromagnetic contribution due to the large value of the topological susceptibility of the vacuum.



\section{Acknowledgments}
I am grateful to O.\,V. Teryaev for interest to this work and stimulating discussions.  


\end{document}